\pdfoutput=1
\documentclass[12pt]{article}
\usepackage{graphicx}
\usepackage{amsmath, amssymb}

\renewcommand{\bar}[1]{\overline{#1}}

\def\Dslash{\raise.15ex\hbox{/}\kern-.7em D}
\def\Pslash{\raise.15ex\hbox{/}\kern-.7em P}

\begin {document}
\begin{flushright}
{\small
SLAC--PUB--12805\\
September 2007}
\end{flushright}

\vspace{20pt}

\centerline{\LARGE \bf Novel QCD Effects}
\vspace{5pt}
\centerline{\LARGE \bf from Initial and Final State Interactions}

\vspace{15pt}

\centerline{\bf {
Stanley J. Brodsky\footnote{Electronic address:
sjbth@slac.stanford.edu}$^{a}$ }}

\vspace{10pt}

{\centerline {$^{a}$Stanford Linear Accelerator Center, 
Stanford University, Stanford, CA 94309, USA}

 \vspace{15pt}

\begin{abstract}

Initial-state and
final-state interactions, which are conventionally neglected in the parton model, have a profound effect in QCD hard-scattering reactions.  These effects,  which arise from gluon exchange between the active and spectator quarks, cause
leading-twist single-spin asymmetries, diffractive deep inelastic scattering, diffractive hard hadronic reactions, and the breakdown of
the Lam-Tung relation in Drell-Yan reactions.  Diffractive deep inelastic scattering also leads to nuclear shadowing and non-universal antishadowing of nuclear structure functions through multiple scattering reactions in the nuclear target. Factorization-breaking effects are particularly important for hard hadron interactions since both initial-state and final-state interactions appear.  Related factorization breaking effects can also appear in exclusive electroproduction reactions and in deeply virtual Compton scattering. 
None of the effects of initial-state and final-state interactions are incorporated in
the light-front wavefunctions of the target hadron computed in isolation. 

\end{abstract}

\vspace{5pt}

\begin{center}
{\it Invited talk presented at  \\
Workshop on Exclusive Reactions at High Momentum Transfer\\
21-24 May 2007  \\
Newport News, Virginia\\
 }
\end{center}

\vfill

\newpage

\parindent=1.5pc
\baselineskip=16pt

\setcounter{footnote}{0}


\vspace{10pt}

\section{Introduction}

Deep inelastic lepton scattering provides  a direct window to the fundamental quark and gluon structure of nucleons and the nucleus.  
 
In the conventional description of of deep inelastic lepton scattering, the final-state interactions of the struck quark can be systematically neglected at leading order in $1/Q^2$. This intuitive picture, which is based on the quark-parton model, is reinforced by the argument that the Wilson line which describes the gauge interactions of the outgoing colored quark current can be set to unity simply by choosing the light cone gauge $A^+ = A^0 + A^3 =0.$  In this intuitive picture, the leading-twist structure functions of the target hadron or nucleus can be computed as a probability distribution in $x$ defined from the square of its light-front wavefunctions.  The Bjorken variable $x_{bj}$ can be identified at leading twist with the light-cone fraction $x^+ = {k^+/ P^+} = {(k^0+ k^3) /(P^0 + P^3)}$ of the struck quark.

Surprising, this simple picture is not actually correct in QCD. The effects of final-state gluonic interactions of the struck quark cannot be neglected in any gauge even at very high $Q^2.$  
There are a number of areas of phenomenology where the effects of final-state interactions become manifest: 

(1) in the case of diffractive deep inelastic scattering (DDIS), experiments at HERA~\cite{Derrick:1993xh}, have shown the proton remains intact and separated by a large rapidity in nearly 15\% of the DIS events. This remarkable phenomena can be understood as due to the exchange of gluons in the final state which neutralize the color separation~\cite{Brodsky:2002ue}, an effect which persists at high $s$ and $Q^2$ in any gauge.  This is illustrated in fig.~\ref{fig26}.
The net effect of the FSI is in fact not even unitary, so that standard interpretation of structure functions as probability distributions is not accurate.      The Wilson line can also be interpreted as the final-state phase induced by a noncausal gauge induced in the DIS reaction. Thus the structure functions measured in DIS cannot be computed solely from the wavefunctions of a hadron in isolation.  This picture also contradicts models based on an intrinsic pomeron component of the proton.

\begin{figure}[htb]
\centering
\includegraphics[angle=0,width=1.0\textwidth]{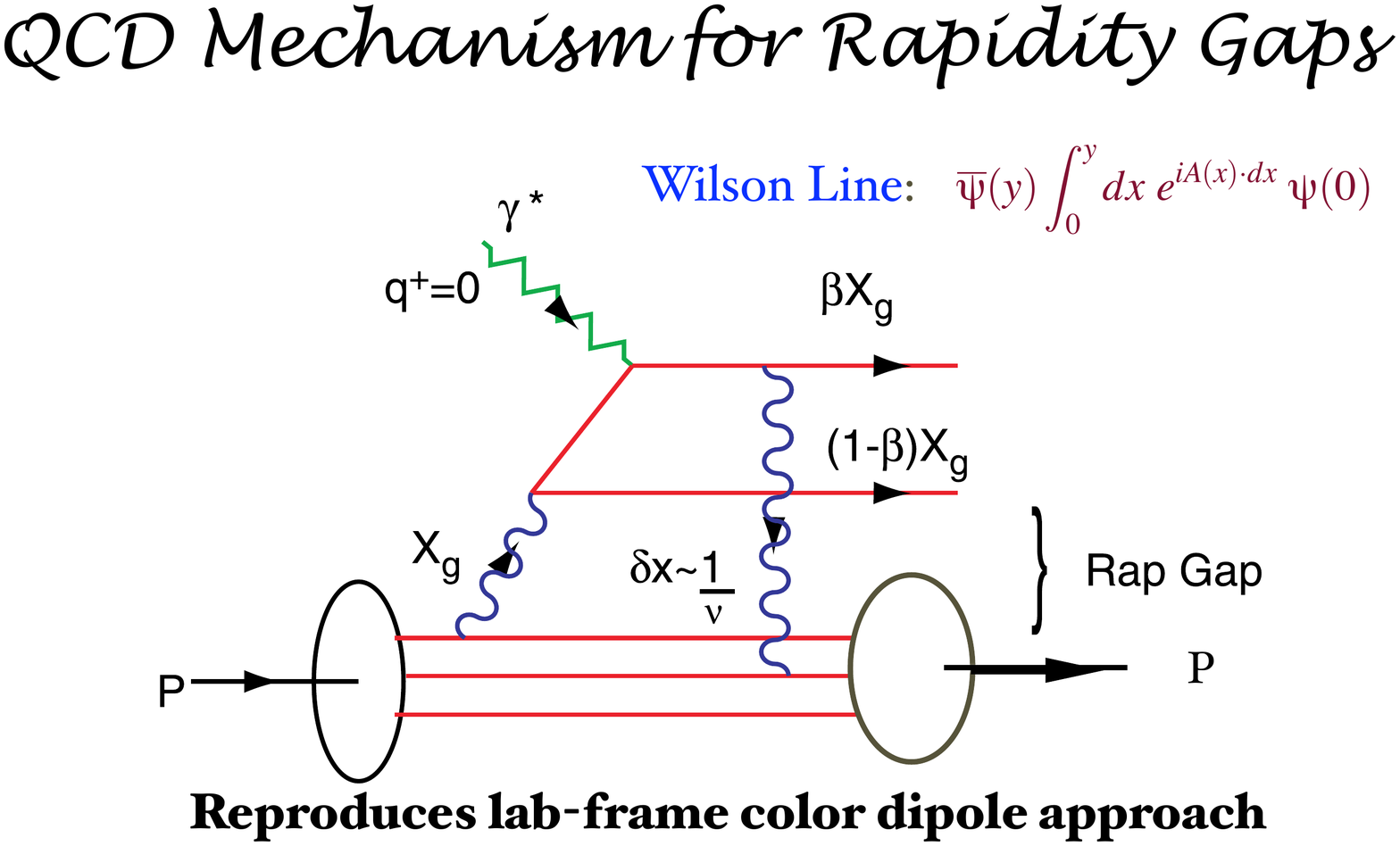}
\caption{Final-state interactions in QCD lead to diffractive deep inelastic scattering 
$ \ell p \to \ell^\prime p^\prime +  X$ at leading twist.} \label{fig26}
\end{figure}

(2) A new understanding of nuclear shadowing and antishadowing has emerged based on the presence of multi-step coherent reactions involving
leading twist diffractive reactions~\cite{Brodsky:1989qz,Brodsky:2004qa} as illustrated in fig. \ref{fig41}.  Thus the nuclear shadowing of structure functions is a consequence of
the lepton-nucleus collision; it is not an intrinsic property of the nuclear wavefunction. The same analysis shows that antishadowing is {\it
not universal}, but it depends in detail on the flavor of the quark or antiquark constituent~\cite{Brodsky:2004qa}.

\begin{figure}[htb]
\centering
\includegraphics[angle=0,width=1.0\textwidth]{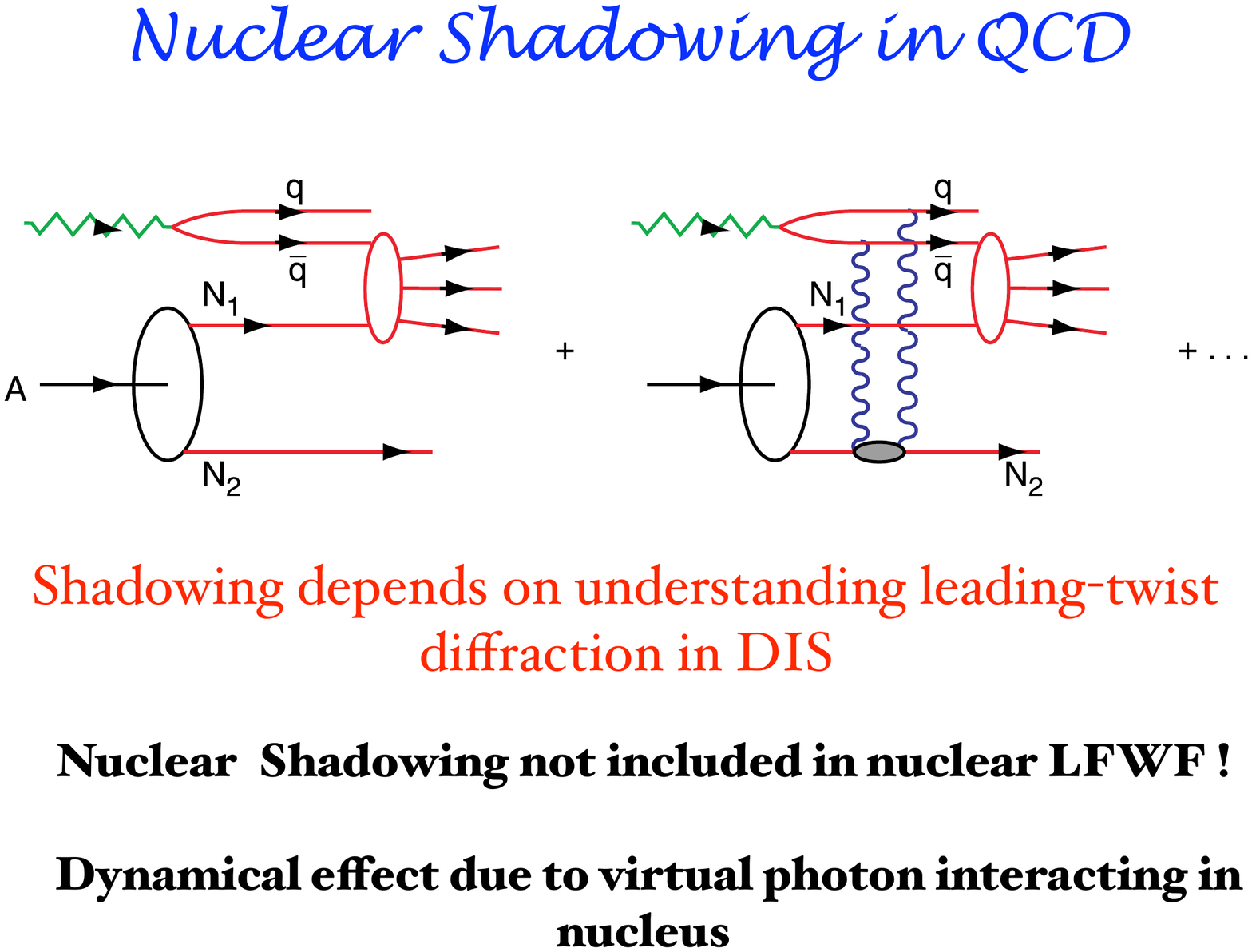}
\caption{Relation of nuclear shadowing of structure functions to leading-twist diffractive deep inelastic scattering.} \label{fig41}
\end{figure}

(3) The final-state interactions illustrated in fig.~\ref{fig12} which cause DDIS also produce the pseudo-T-odd  
$i ~\vec S_p \cdot \vec p_H \times \vec q ~$   Sivers correlation between the production plane of the hadron or jet produced in semi-inclusive deep inelastic scattering (DIS) and the target nucleon spin.
These interactions produce the Sivers
effect at leading twist~\cite{Brodsky:2002cx} with different signs in semi-inclusive deep inelastic scattering and the Drell-Yan
reaction~\cite{Collins:2002kn}.  Double initial-state interactions~\cite{Boer:2002ju} also produce anomalous angular effects, including the
breakdown of the Lam-Tung relation~\cite{Lam:1980uc} in the Drell-Yan process. 
As recently noted by Collins and Qiu~\cite{Collins:2007nk}, the traditional factorization formalism of perturbative QCD for high transverse
momentum hadron production in hadron collisions also fails in detail because of initial- and final-state gluonic interactions. 

\begin{figure}[htb]
\centering
\includegraphics[angle=0,width=1.0\textwidth]{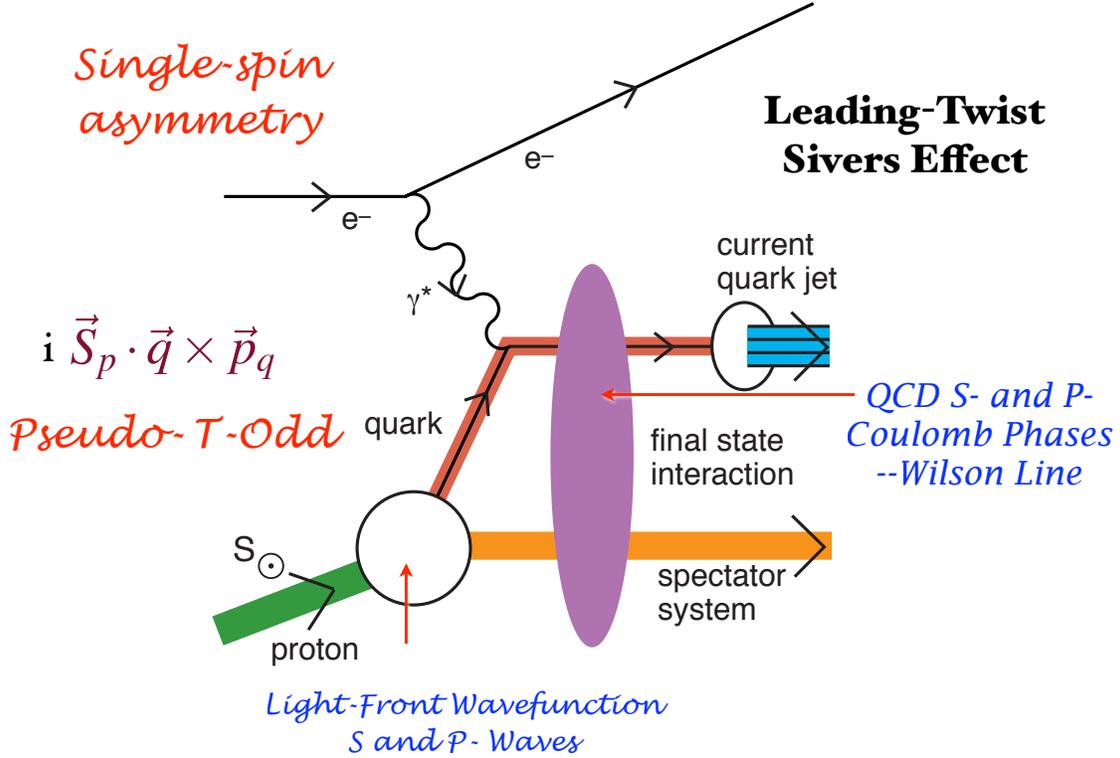}
\caption{Final-state interactions in QCD and the physics of the leading-twist Sivers single-spin asymmetry in semi-inclusive deep inelastic
lepton-proton scattering.} \label{fig12}
\end{figure}

These examples of unconventional wisdom highlight the need for a fundamental understanding the dynamics of hadrons in QCD at the amplitude
level. This is essential for understanding phenomena such as the quantum mechanics of hadron formation and the origins of diffractive phenomena, as well and single-spin asymmetries.

\section{Diffractive Deep Inelastic Scattering}
A remarkable feature of deep inelastic lepton-proton scattering at HERA is that up to 15\% of the events are
diffractive~\cite{Adloff:1997sc,Breitweg:1998gc}: the target proton remains intact, and there is a large rapidity gap between the proton and the
other hadrons in the final state.  These diffractive deep inelastic scattering (DDIS) events can be understood most simply from the perspective
of the color-dipole model: the $q \bar q$ Fock state of the high-energy virtual photon diffractively dissociates into a diffractive dijet
system.  The exchange of multiple gluons between  the color dipole of the $q \bar q$ and the quarks of the target proton neutralizes the color
separation and leads to the diffractive final state.  The same multiple gluon exchange also controls diffractive vector meson electroproduction
at large photon virtuality \cite{Brodsky:1994kf}.  This observation presents a paradox: if one chooses the conventional parton model frame where
the photon light-front momentum is negative $q+ = q^0 + q^z  < 0$, the virtual photon interacts with a quark constituent with light-cone
momentum fraction $x = {k^+/p^+} = x_{bj}.$  Furthermore, the gauge link associated with the struck quark (the Wilson line) becomes unity in
light-cone gauge $A^+=0$. Thus the struck ``current" quark apparently experiences no final-state interactions. Since the light-front
wavefunctions $\psi_n(x_i,k_{\perp i})$ of a stable hadron are real, it appears impossible to generate the required imaginary phase associated
with pomeron exchange, let alone large rapidity gaps.

This paradox was resolved by Hoyer, Marchal,  Peigne, Sannino and myself \cite{Brodsky:2002ue}.  Consider the case where the virtual photon
interacts with a strange quark---the $s \bar s$ pair is assumed to be produced in the target by gluon splitting.  In the case of Feynman gauge,
the struck $s$ quark continues to interact in the final state via gluon exchange as described by the Wilson line. The final-state interactions
occur at a light-cone time $\Delta\tau \simeq 1/\nu$ shortly after the virtual photon interacts with the struck quark. When one integrates over
the nearly-on-shell intermediate state, the amplitude acquires an imaginary part. Thus the rescattering of the quark produces a separated
color-singlet $s \bar s$ and an imaginary phase. In the case of the light-cone gauge $A^+ = \eta \cdot A =0$, one must also consider the
final-state interactions of the (unstruck) $\bar s$ quark. The gluon propagator in light-cone gauge $d_{LC}^{\mu\nu}(k) = (i/k^2+ i
\epsilon)\left[-g^{\mu\nu}+\left(\eta^\mu k^\nu+ k^\mu\eta^\nu / \eta\cdot k\right)\right] $ is singular at $k^+ = \eta\cdot k = 0.$ The
momentum of the exchanged gluon $k^+$ is of ${ \cal O}{(1/\nu)}$; thus rescattering contributes at leading twist even in light-cone gauge.  This is illustrated in fig.~\ref{fig28}.  The
net result is  gauge invariant and is identical to the color dipole model calculation. The calculation of the rescattering effects on DIS in
Feynman and light-cone gauge through three loops is given in detail for an Abelian model in reference~\cite{Brodsky:2002ue}.  The result shows
that the rescattering corrections reduce the magnitude of the DIS cross section in analogy to nuclear shadowing.

\begin{figure}[htb]
\centering
\includegraphics[angle=0,width=1.0\textwidth]{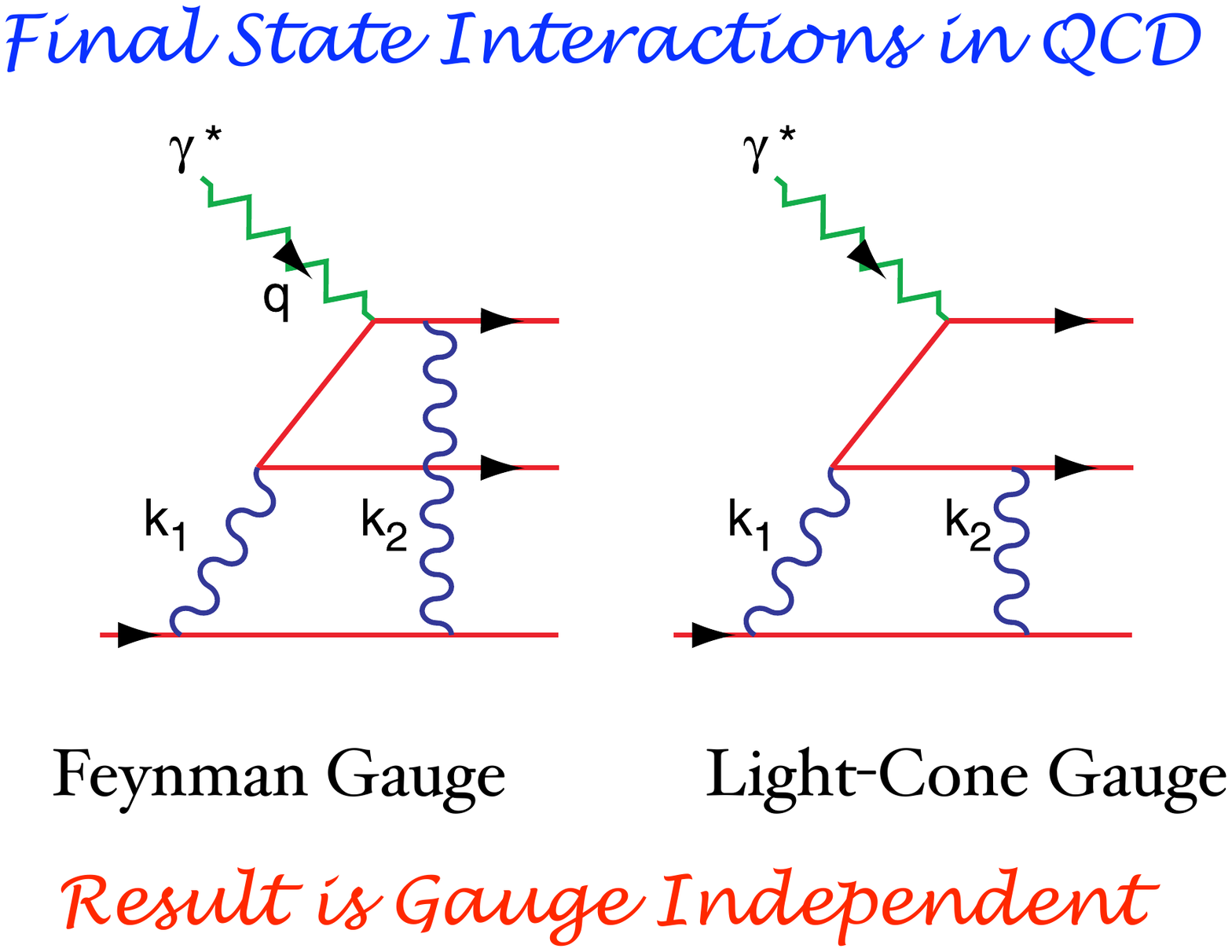}
\caption{Final-state interactions in QCD are nonzero even in light-cone gauge.} \label{fig28}
\end{figure}

A new understanding of the role of final-state interactions in deep inelastic scattering has thus emerged. The multiple scattering of the struck
parton via instantaneous interactions in the target generates dominantly imaginary diffractive amplitudes, giving rise to an effective ``hard
pomeron" exchange.  The presence of a rapidity gap between the target and diffractive system requires that the target remnant emerges in a
color-singlet state; this is made possible in any gauge by the soft rescattering.  The resulting diffractive contributions leave the target
intact  and do not resolve its quark structure; thus there are contributions to the DIS structure functions which cannot be interpreted as
parton probabilities~\cite{Brodsky:2002ue}; the leading-twist contribution to DIS  from rescattering of a quark in the target is a coherent
effect which is not included in the light-front wave functions computed in isolation. One can augment the light-front wave functions with a
gauge link corresponding to an external field created by the virtual photon $q \bar q$ pair current~\cite{Belitsky:2002sm,Collins:2004nx}.  Such
a gauge link is process dependent~\cite{Collins:2002kn}, so the resulting augmented LFWFs are not universal
\cite{Brodsky:2002ue,Belitsky:2002sm,Collins:2003fm}.   We also note that the shadowing of nuclear structure functions is due to the destructive
interference between multi-nucleon amplitudes involving diffractive DIS and on-shell intermediate states with a complex phase. In contrast, the
wave function of a stable target is strictly real since it does not have on-energy-shell intermediate state configurations.  The physics of
rescattering and shadowing is thus not included in the nuclear light-front wave functions, and a probabilistic interpretation of the nuclear DIS
cross section is precluded.

Rikard Enberg, Paul Hoyer, Gunnar Ingelman and I~\cite{Brodsky:2004hi} have shown that the quark structure function of the effective hard
pomeron has the same form as the quark contribution of the gluon structure function. The hard pomeron is not an intrinsic part of the proton;
rather it must be considered as a dynamical effect of the lepton-proton interaction. Our QCD-based picture also applies to diffraction in
hadron-initiated processes. The rescattering is different in virtual photon- and hadron-induced processes due to the different color
environment, which accounts for the  observed non-universality of diffractive parton distributions. This framework also provides a theoretical
basis for the phenomenologically successful Soft Color Interaction (SCI) model~\cite{Edin:1995gi} which includes rescattering effects and thus
generates a variety of final states with rapidity gaps.

\section{Shadowing and Antishadowing of Nuclear Structure Functions}

One of the novel features of QCD involving nuclei is the {\it antishadowing} of the nuclear structure functions which is observed in deep
inelastic lepton scattering and other hard processes. Empirically, one finds $R_A(x,Q^2) \equiv  \left(F_{2A}(x,Q^2)/ (A/2) F_{d}(x,Q^2)\right)
> 1 $ in the domain $0.1 < x < 0.2$; {\em i.e.}, the measured nuclear structure function (referenced to the deuteron) is larger than than the
scattering on a set of $A$ independent nucleons.

The shadowing of the nuclear structure functions: $R_A(x,Q^2) < 1 $ at small $x < 0.1 $ can be readily understood in terms of the Gribov-Glauber
theory.  Consider a two-step process in the nuclear target rest frame. The incoming $q \bar q$ dipole first interacts diffractively $\gamma^* +
N_1 \to (q \bar q) N_1$ on nucleon $N_1$ leaving it intact.  This is the leading-twist diffractive deep inelastic scattering  (DDIS) process
which has been measured at HERA to constitute approximately 10\% of the DIS cross section at high energies.  The $q \bar q$ state then interacts
inelastically on a downstream nucleon $N_2:$ $(q \bar q) N_2 \to X$. The phase of the pomeron-dominated DDIS amplitude is close to imaginary,
and the Glauber cut provides another phase $i$, so that the two-step process has opposite  phase and  destructively interferes with the one-step
DIS process $\gamma^* + N_2 \to X$ where $N_1$ acts as an unscattered spectator. The one-step and-two step amplitudes can coherently interfere as
long as the momentum transfer to the nucleon $N_1$ is sufficiently small that it remains in the nuclear target;  {\em i.e.}, the Ioffe
length~\cite{Ioffe:1969kf} $L_I = { 2 M \nu/ Q^2} $ is large compared to the inter-nucleon separation. In effect, the flux reaching the interior
nucleons is diminished, thus reducing the number of effective nucleons and $R_A(x,Q^2) < 1.$

There are also leading-twist diffractive contributions $\gamma^* N_1 \to (q \bar q) N_1$  arising from Reggeon exchanges in the
$t$-channel~\cite{Brodsky:1989qz}.  For example, isospin--non-singlet $C=+$ Reggeons contribute to the difference of proton and neutron
structure functions, giving the characteristic Kuti-Weisskopf $F_{2p} - F_{2n} \sim x^{1-\alpha_R(0)} \sim x^{0.5}$ behavior at small $x$. The
$x$ dependence of the structure functions reflects the Regge behavior $\nu^{\alpha_R(0)} $ of the virtual Compton amplitude at fixed $Q^2$ and
$t=0.$ The phase of the diffractive amplitude is determined by analyticity and crossing to be proportional to $-1+ i$ for $\alpha_R=0.5,$ which
together with the phase from the Glauber cut, leads to {\it constructive} interference of the diffractive and nondiffractive multi-step nuclear
amplitudes. Furthermore, because of its $x$ dependence, the nuclear structure function is enhanced precisely in the domain $0.1 < x <0.2$ where
antishadowing is empirically observed.  The strength of the Reggeon amplitudes is fixed by the fits to the nucleon structure functions, so there
is little model dependence.

As noted above, the Bjorken-scaling diffractive contribution to DIS arises from the rescattering of the struck quark after it is struck  (in the
parton model frame $q^+ \le 0$), an effect induced by the Wilson line connecting the currents. Thus one cannot attribute DDIS to the physics of
the target nucleon computed in isolation~\cite{Brodsky:2002ue}.  Similarly, since shadowing and antishadowing arise from the physics of
diffraction, we cannot attribute these phenomena to the structure of the nucleus itself: shadowing and antishadowing arise because of the
$\gamma^* A$ collision and the history of the $q \bar q$ dipole as it propagates through the nucleus.

Ivan Schmidt, Jian-Jun Yang, and I~\cite{Brodsky:2004qa} have extended the Glauber analysis to the shadowing and antishadowing of all of the
electroweak structure functions. Quarks of different flavors  will couple to different Reggeons; this leads to the remarkable prediction that
nuclear antishadowing is not universal; it depends on the quantum numbers of the struck quark. This picture implies substantially different
antishadowing for charged and neutral current reactions, thus affecting the extraction of the weak-mixing angle $\theta_W$.  We find that part
of the anomalous NuTeV result~\cite{Zeller:2001hh} for $\theta_W$ could be due to the non-universality of nuclear antishadowing for charged and
neutral currents. Detailed measurements of the nuclear dependence of individual quark structure functions are thus needed to establish the
distinctive phenomenology of shadowing and antishadowing and to make the NuTeV results definitive. Schmidt, Yang, and I have also identified
contributions to the nuclear multi-step reactions which arise from odderon exchange and hidden color degrees of freedom in the nuclear
wavefunction. There are other ways in which this new view of antishadowing can be tested;  antishadowing can also depend on the target and beam
polarization.

\section{ Single-Spin Asymmetries from Initial- and Final-State
Interactions}

Among the most interesting polarization effects are single-spin azimuthal asymmetries  in semi-inclusive deep inelastic scattering, representing
the correlation of the spin of the proton target and the virtual photon to hadron production plane: $\vec S_p \cdot \vec q \times \vec p_H$.
Such asymmetries are time-reversal odd, but they can arise in QCD through phase differences in different spin amplitudes. In fact, final-state
interactions from gluon exchange between the outgoing quarks and the target spectator system lead to single-spin asymmetries in semi-inclusive
deep inelastic lepton-proton scattering  which  are not power-law suppressed at large photon virtuality $Q^2$ at fixed
$x_{bj}$~\cite{Brodsky:2002cx}.  In contrast to the SSAs arising from transversity and the Collins fragmentation function, the fragmentation of
the quark into hadrons is not necessary; one predicts a correlation with the production plane of the quark jet itself. Physically, the
final-state interaction phase arises as the infrared-finite difference of QCD Coulomb phases for hadron wave functions with differing orbital
angular momentum.  This is illustrated in  fig.~\ref{fig12}. The same proton matrix element which determines the spin-orbit correlation $\vec S \cdot \vec L$ also
produces the anomalous magnetic moment of the proton, the Pauli form factor, and the generalized parton distribution $E$ which is measured in
deeply virtual Compton scattering. Thus the contribution of each quark current to the SSA is proportional to the contribution $\kappa_{q/p}$ of
that quark to the proton target's anomalous magnetic moment $\kappa_p = \sum_q e_q \kappa_{q/p}$ ~\cite{Brodsky:2002cx,Burkardt:2004vm}.  
The
HERMES collaboration has recently measured the SSA in pion and kaon electroproduction using transverse target polarization~\cite{Airapetian:2004tw}. The
Sivers and Collins effects can be separated using planar correlations; both contributions are observed to contribute, with values not in
disagreement with theory expectations ~\cite{Airapetian:2004tw,Avakian:2004qt}.  The larger Sivers effect seem for $K^+$ production compared to $\pi+$ production at small $x_{bj}$ suggests a role for polarized sea quarks.

\begin{figure}[htb]
\centering
\includegraphics[angle=0,width=1.0\textwidth]{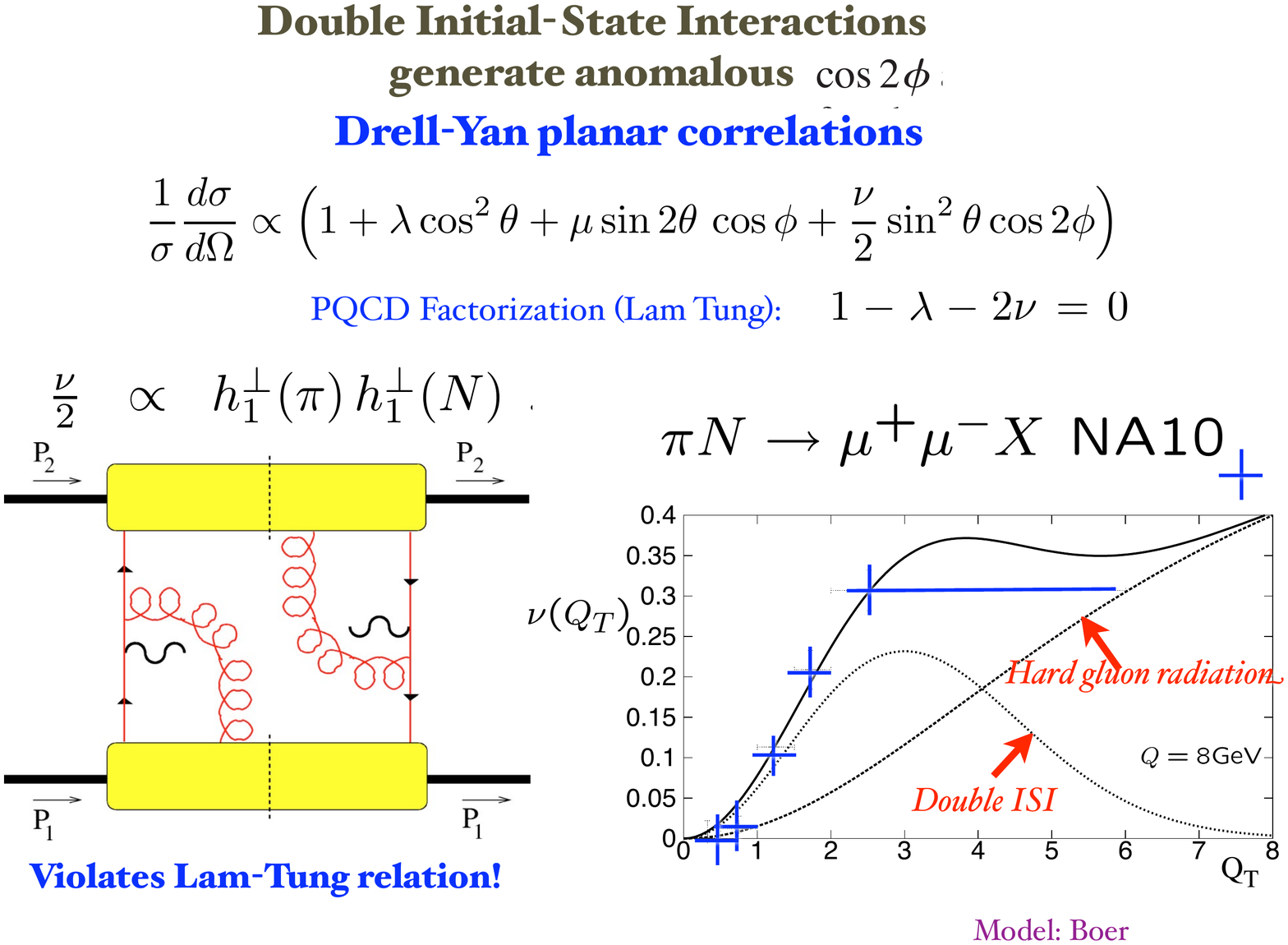}
\caption{Double initial-state interactions in QCD violate the Lam-Tung relation for Drell-Yan massive lepton pair production.} \label{fig18}
\end{figure}

The final-state interaction mechanism provides an appealing physical explanation within QCD of single-spin asymmetries.  Remarkably, the same
matrix element which determines the spin-orbit correlation $\vec S \cdot \vec L$ also produces the anomalous magnetic moment of the proton, the
Pauli form factor, and the generalized parton distribution $E$ which is measured in deeply virtual Compton scattering.  Physically, the
final-state interaction phase arises as the infrared-finite difference of QCD Coulomb phases for hadron wave functions with differing orbital
angular momentum.  An elegant discussion of the Sivers effect including its sign has been given by Burkardt~\cite{Burkardt:2004vm}. As shown
recently by Gardner and myself~\cite{Brodsky:2006ha}, one can also use the Sivers effect to study the orbital angular momentum of  gluons by
tagging a gluon jet in semi-inclusive DIS. In this case, the final-state interactions are enhanced by the large color charge of the gluons.

The final-state interaction effects can also be identified with the gauge link which is present in the gauge-invariant definition of parton
distributions~\cite{Collins:2004nx}.  Even in light-cone gauge, a transverse gauge link is required which reflects the external conditions of electroproduction. Thus the
parton amplitudes need to be augmented by an additional eikonal factor incorporating the final-state interaction and its
phase~\cite{Ji:2002aa,Belitsky:2002sm}. This procedure allows one to formally define transverse momentum dependent parton distribution
functions which contain the effect of the QCD final-state interactions.   However, the physics of final state interactions is not contained in the wavefunction of a hadron in isolation.

\section{The Exclusive Sivers Effect} 

It would also be interesting to study the Sivers effect in exclusive electroproduction reactions. For example, there should be a  $i \vec S_p \cdot \vec q \times \vec p_\pi $ correlation in pion electroproduction
$\gamma^* p_\updownarrow \to \pi^+  n.$   This could be an ideal experiment for the 12 GeV program at JLab.

A central uncertainty in the analysis of B decays is the unknown nature and magnitude of the strong phase.  
It would thus be interesting to make a connection between the final-state hadronic phases  which cause the Sivers effect in exclusive electroproduction and the strong interaction phases which appear in exclusive B decays.  The final-state QCD phase in such hard processes would be expected to be diminished because of color transparency as the momentum transfer squared  $t$ to the meson increases. 
 
\section{The Sivers Effect in General Inclusive Reactions}

A related analysis also predicts that the initial-state
interactions from gluon exchange between the incoming quark and the target spectator system lead to leading-twist single-spin asymmetries in the
Drell-Yan process $H_1 H_2^\updownarrow \to \ell^+ \ell^- X$ ~\cite{Collins:2002kn,Brodsky:2002rv}.  The SSA in the Drell-Yan process is the
same as that obtained in SIDIS, with the appropriate identification of variables, but with the opposite sign. There is no Sivers effect in
charged-current reactions since the $W$ only couples to left-handed quarks~\cite{Brodsky:2002pr}.

If both the quark and antiquark in the initial state of the Drell-Yan subprocess $q \bar q \to  \mu^+ \mu^-$ interact with the spectators of the
other incident hadron, one finds a breakdown of the Lam-Tung relation, which was formerly believed to be a general prediction of leading-twist
QCD. These double initial-state interactions also lead to a $\cos 2 \phi$ planar correlation in unpolarized Drell-Yan
reactions~\cite{Boer:2002ju}. More generally one must consider subprocesses involving initial-state gluons such as $n g q \bar q \to \ell \bar
\ell$  as well as subprocesses with extra final-state gluons.  This is illustrated in fig. ~\ref{fig18}.
\begin{figure}[htb]
\centering
\includegraphics[angle=0,width=1.0\textwidth]{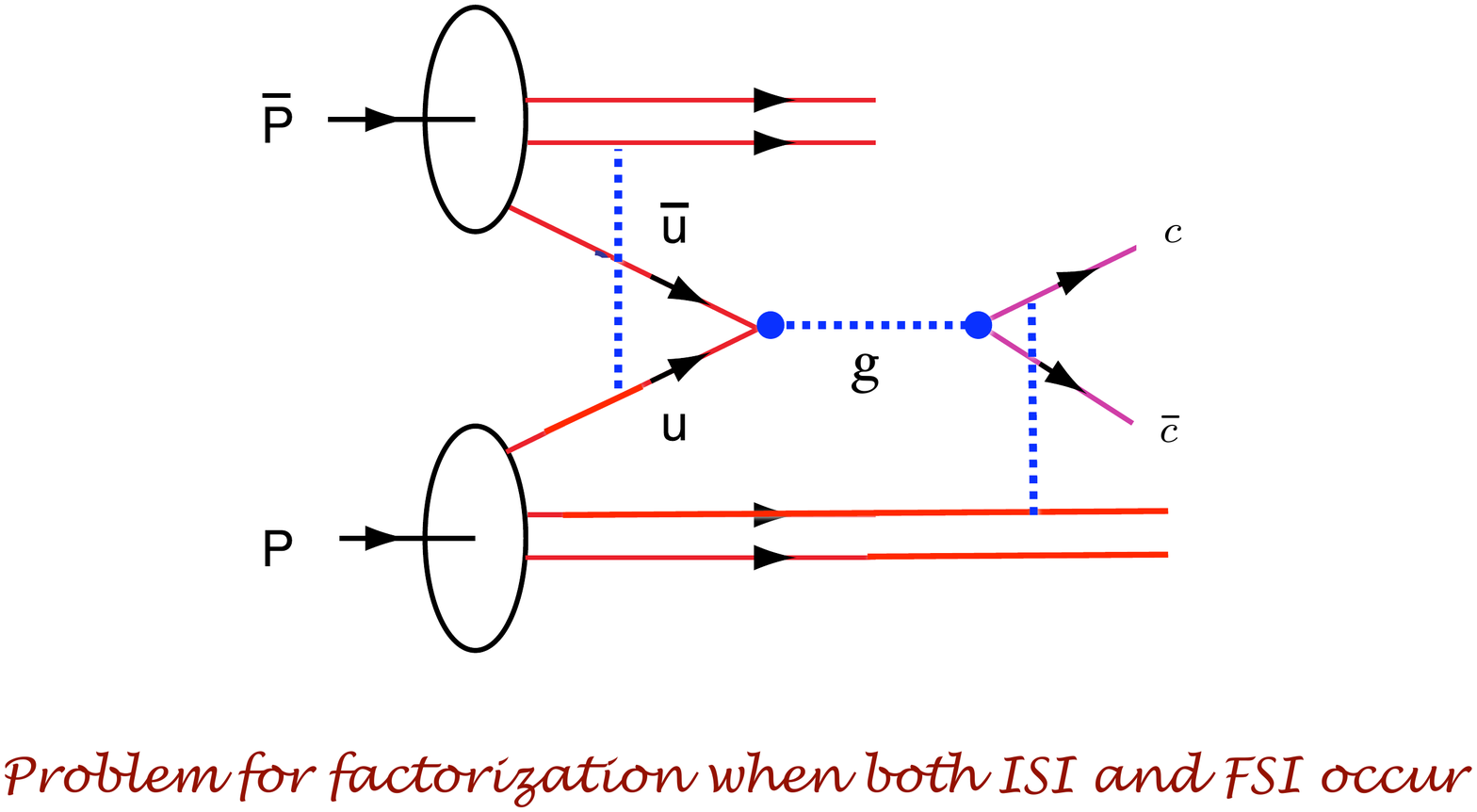}
\caption{Initial-state and final-state interactions in QCD both contribute to massive heavy quark production.} \label{fig20}
\end{figure}

The situation becomes more complicated in the case of hard hadron interactions where both initial and final state interactions are present.  An example involving heavy quark production is shown in  fig. ~\ref{fig20}.   As noted by Collins and Qiu~\cite{Collins:2007nk} the combination of such effects endanger the standard arguments for factorization in general hadroproduction processes.   In addition, the final-state interactions which produce diffractive deep inelastic scattering and the Sivers effect in leptoproduction at leading twist will also affect the intermediate quark line in the virtual Compton amplitude, thus correcting the handbag approximation to DVCS.

\section{Summary}

Initial- and
final-state interactions from gluon-exchange, which are neglected in the parton model, have a profound effect in QCD hard-scattering reactions.  These effects cause
leading-twist single-spin asymmetries, diffractive deep inelastic scattering, diffractive hard hadronic reactions, and the breakdown of
the Lam Tung relation in Drell-Yan reactions. Diffractive deep inelastic scattering leads to nuclear shadowing and non-universal antishadowing.  Related effects can appear in exclusive electroproduction reactions and in deeply virtual Compton scattering.   None of the effects of initial or final state interactions are incorporated in
the light-front wavefunctions of the target hadron computed in isolation.

\vspace{10pt}

\noindent{\bf Acknowledgments} 

\vspace{5pt}

The results in this talk are based on collaborations with Daniel Boer, Rikard Enberg, Susan Gardner, Paul Hoyer, Dae Sung Hwang, Gunnar Ingelman, Hung Jung Lu,  Nils Marchal,  Stephane Peigne,  Francesco Sannino,  and Ivan Schmidt.
This work was supported in
part by the Department of Energy, contract No. DE-AC02-76SF00515.

\end{document}